BOIIARINOVA Y.E.,KALINOVSKIY Y.A., LANDE D.V.

# APPLICATION OF HYPERCOMPLEX NUMERICAL SYSTEMS IN THE DYNAMIC NETWORK MODEL

## 1. INTRODUCTION

In recent years, the direction of the study of networks in which connections correspond to the mutual impact of nodes has been developed. Many works have been devoted to the study of such complex networks [1], but most often they relate to the distribution of one type of activity (inpact). At the same time, there are studies that offer to consider several characteristics of the impact [2]. In particular, various mathematical models are developed and studied: models with thresholds, models of independent cascades, models of epidemics, models of Markov processes, etc. [3]

In this paper, it is proposed to use hypercomplex number systems, which are a mathematical apparatus that allows you to model some network problems and solve them at a new level [4,5].

The purpose of this work is to model the propagation of several characteristics of the effects on the nodes of a complex network using the apparatus of hypercomplex number systems (HNS).

## 2. HYPERCOMPLEX MODEL OF DISTRIBUTION OF SEVERAL TYPES OF ACTIVITY

У цій роботі розглядаються мережі, де ребра відповідають взаємним впливам вузлів. Зазвичай розглядаються динамічні мережеві моделі, засновані на розповсюдженні одного виду активності. Запропонована модель, навпаки, базується на мережах, в яких передбачено багато видів активності. Нехай складна мережа – це граф, який складається з $N$ вузлів, між якими є зв'язки між вузлами $i$ та $j$ (ребра) з вагою $L_{i,j}$, $i, j = 1 \dots N$. Якщо ребро між вузлами відсутнє, то враховуємо вагу зв'язку $L_{i,j} = 0$.

В загальному вигляді така складна система виглядає, як зображено на рис.1

In this paper, we consider networks where the edges correspond to the mutual impact of the nodes. Dynamic network models based on the distribution of one type of activity are usually considered. The proposed model is based on networks that provide many types of activity. Let a complex network be a graph consisting of N nodes, between which there are connections between nodes $i$ and $j$ (edges) with weight $L_{i,j}$, $i, j = 1 \dots N$. If the edge between the nodes is missing, then consider the weight of the connection $L_{i,j} = 0$

In general, such a complex system looks like shown in Fig.1

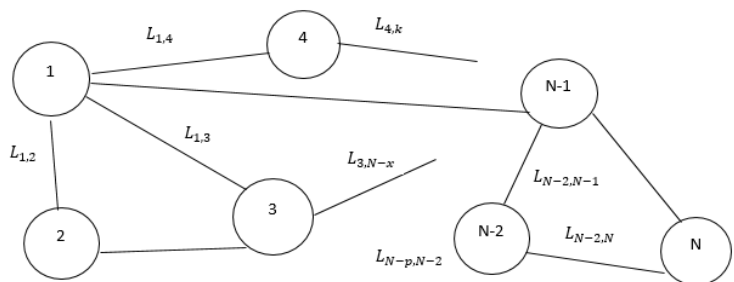

*Figure 1. General view of a complex system*

In the proposed model, we consider the possibility that each node has several properties that can be described by a hypercomplex numerical system of dimension T in the general case.

The impact of the properties of one node on another can occur according to the Kelly table of some HNS. Consider the case when the node has only two properties T = 2. Then the Kelly table consists of four cells ($T^2 = 4$).

| $\Gamma(e,2)$ | $e_1$ | $e_2$ |
|---|---|---|
| $e_1$ | $a_1 e_1 + a_2 e_2$ | $b_1 e_1 + b_2 e_2$ |
| $e_2$ | $b_1 e_1 + b_2 e_2$ | $c_1 e_1 + c_2 e_2$ |

In this Kelly table, all structural constants are real numbers. Due to isomorphic transformations, you can go to the Kelly table as follows:

| $\Gamma_5(f,2)$ | $f_1$ | $f_2$ |
|---|---|---|
| $f_1$ | $f_1$ | $f_2$ |
| $f_2$ | $f_2$ | $c_1 c_2 f_1 + c_2 f_2$ |

The system $\Gamma_5(f,2)$ is isomorphic to one of the "classical" systems: complex $C(e,2)$, dual $D(e,2)$, double $W(e,2)$. Thus, considering the effect of one node on another with two properties, you can use HNS $\Gamma_5(f,2)$.

## 3. APPLICATION OF THE HYPERCOMPLEX COMPUTING SOFTWARE

The main difficulty of performing analytical operations with hypercomplex numbers is their multidimensionality and the related fact is the cumbersomeness in the analytical form of expressions from multidimensional hypercomplex numbers. To do this, a package of programs and procedures based on the system of computer algebra Maple [5]. Structurally, the software of hypercomplex calculations (SHC) consists of the following subsystems: algebraic operations in the HNS, manipulation of the HNS and Kelly tables; determination of algebraic characteristics of hypercomplex expressions; storage of frequently used expressions; performing modular operations with hypercomplex expressions; visualization and service. This structure and composition of SHC in the Maple environment can greatly simplify the process of creating software for mathematical modeling of various scientific and technical problems.

## 4. RESULTS

The simplest example, when the system consists of three nodes, in each node there are two properties that correspond to the HNS 2nd dimension.

The system of equations of this model is easy to build with the help of SHC.

Consider the case of a network consisting of three nodes.

$$
\begin{aligned}
&a_{1,1}\,a_{2,1} + a_{1,2}\,a_{2,2}\,p = l_{1,2} && a_{1,1}\,a_{2,2} + a_{1,2}\,a_{2,1} + a_{1,2}\,a_{2,2}\,q = 0 \\
&a_{1,1}\,a_{3,1} + a_{1,2}\,a_{3,2}\,p = l_{1,3} && a_{1,1}\,a_{2,2} + a_{1,2}\,a_{2,1} + a_{1,2}\,a_{2,2}\,q = 0 \\
&a_{2,1}\,a_{3,1} + a_{2,2}\,a_{3,2}\,p = l_{2,3} && a_{1,1}\,a_{3,2} + a_{1,2}\,a_{3,1} + a_{1,2}\,a_{3,2}\,q = 0 \qquad\qquad (1) \\
&a_{2,1}\,a_{3,2} + a_{2,2}\,a_{3,1} + a_{2,2}\,a_{3,2}\,q = 0
\end{aligned}
$$

This system consists of 6 equations, unknown variable in which there will be 6 - $a_{ij}$. System (1) has 2 solutions. The first solution has zero coefficients $a_{1,2} = a_{2,2} = a_{3,2} = 0$, ie go to the network with real parameters. The second solution

$$
a_{1,1} = -\frac{\sqrt{\dfrac{l_{1,2}\,l_{2,3}}{4\,p\,l_{1,3} + l_{1,3}\,q^2}}\,l}{l_{2,3}} \quad a_{1,2} = \frac{2\sqrt{\dfrac{l_{1,2}\,l_{2,3}}{4\,p\,l_{1,3} + l_{1,3}\,q^2}}\,l_{1,3}}{l_{2,3}} \quad a_{2,1} = -\sqrt{\dfrac{l_{1,2}\,l_{2,3}}{4\,p\,l_{1,3} + l_{1,3}\,q^2}}\,q
$$

$$
a_{2,2} = 2\sqrt{\dfrac{l_{1,2}\,l_{2,3}}{4\,p\,l_{1,3} + l_{1,3}\,q^2}} \quad a_{3,1} = -\frac{q\,l_{2,3}}{\sqrt{\dfrac{l_{1,2}\,l_{2,3}}{4\,p\,l_{1,3} + l_{1,3}\,q^2}}}\,(q^2 + \quad a_{3,2} = \frac{2\,l_{2,3}}{\sqrt{\dfrac{l_{1,2}\,l_{2,3}}{4\,p\,l_{1,3} + l_{1,3}\,q^2}}}\,(q^2 +
$$

This solution depends on the parameters $p$ and $q$ which are the structural constants of the HNS. That is, by selecting these parameters, you can select the type of required HNS. Suppose, $p = 1, q = 0.5, l_{1,2} = 7, l_{2,3} = 8, l_{1,3} = 2$ then, the corresponding coefficients of impact $a_{1,1} = -0.32$, $a_{1,2} = 0.64$, $a_{2,1} = -1.28$, $a_{2,2} = 5.13$, $a_{3,1} = -0.36$, $a_{3,2} = 1.47$.

When increasing the number of nodes, the relationship between the number of equations in the system and the number of unknowns variables must be taken into account

## CONCLUSIONS

The paper considers examples of building a complex network. In the presented network model, the edges correspond to the mutual impact of the nodes, each of which has several types of activity. Two types of activity corresponding to hypercomplex numerical systems of the second dimension that are isomorphic to systems $\Gamma_5(f, 2)$ were considered. The modeling of the network operation with the hypercomplex number systems is constructed with the SHC means. Using of SHC it is possible to create a complex network of any dimension with several types of activity. The study of systems with more activity depends on the power of the computer system on which the SHC can be installed.